\documentclass[twocolumn,preprintnumbers,superscriptaddress,nofootinbib,aps,prd]{revtex4-1}

\pdfoutput=1

\usepackage[hypertex]{hyperref}

\usepackage{graphicx}
\usepackage{dcolumn}
\usepackage{amssymb}
\usepackage{amsmath}

\usepackage{mathrsfs} 
\usepackage[FIGTOPCAP]{subfigure}
\usepackage{comment}

\usepackage{epsfig}
\usepackage{latexsym}

\newcommand{\lsim}{\lesssim}
\newcommand{\gsim}{\gtrsim}
\newcommand{\ra}{\rightarrow}
\newcommand{\tlab}{\theta_{\rm lab}}
\newcommand{\tcom}{\theta_{\rm com}}

\def\be{\begin{equation}}
\def\ee{\end{equation}}
\def\bea{\begin{eqnarray}}
\def\eea{\end{eqnarray}}

\newcommand{\kev}{{\rm keV}}

\begin{document}
\pagestyle{plain}

\preprint{UCI-HEP-TR-2010-26}
\preprint{FERMILAB-PUB-10-436-T}

\title{Interpreting Dark Matter Direct Detection \\
Independently of the Local Velocity and Density Distribution}

\author{Patrick J. Fox}
\affiliation{Theoretical Physics Department, Fermilab, Batavia, IL 60510}
\affiliation{School of Natural Sciences, Institute for Advanced Study, 
             Princeton, NJ 08540}
\author{Graham D. Kribs}
\affiliation{Theoretical Physics Department, Fermilab, Batavia, IL 60510}
\affiliation{Department of Physics, University of Oregon, Eugene, OR 97403}
\author{Tim M.P. Tait}
\affiliation{Department of Physics and Astronomy, University of California, 
             Irvine, CA 92697}

\begin{abstract}

We demonstrate precisely what particle physics information can be 
extracted from a single direct detection observation of dark matter
while making absolutely no assumptions about the local 
velocity distribution and local density of dark matter.
Our central conclusions follow from a very simple observation:  
the velocity distribution of dark matter is positive definite, 
$f(v) \ge 0$.  We demonstrate the utility of this 
result in several ways.  First, we show a falling deconvoluted 
recoil spectrum (deconvoluted of the nuclear form factor), 
such as from ordinary elastic scattering,
can be ``mocked up'' by any mass of dark matter above
a kinematic minimum.  As an example, we show that dark matter 
much heavier than previously considered can explain the
CoGeNT excess.  Specifically, $m_\chi < m_{\rm Ge}$ can be
in just as good agreement as light dark matter, while
$m_\chi > m_{\rm Ge}$ depends on understanding the sensitivity
of Xenon to dark matter at very low recoil energies,
$E_R \lsim 6$~keVnr.
Second, we show that any rise in the deconvoluted recoil spectrum
represents distinct particle physics information that \emph{cannot}
be faked by an arbitrary 
$f(v)$.
As examples of resulting non-trivial particle physics, 
we show that inelastic dark matter and dark matter with a 
form factor can both yield such a rise.  

\end{abstract}

\maketitle

\section{Introduction}
\label{sec:intro}

What particle physics information can be extracted from a
detection of events at a dark matter direct detection experiment?
This seemingly innocuous question is riddled with subtleties 
and uncertainties.  The main experimental uncertainties relate
to separating signal from background, and translating observed
energy in a detector to nuclear recoil energy.  The main
astrophysical uncertainties are the local properties
of dark matter for which direct detection is sensitive:
the local velocity distribution and the local density.

The average density of dark matter at a galactic radius equal
to the Sun is reasonably well known based on galactic kinematics
(for a recent discussion, see \cite{Pato:2010yq}).
The local density could be quite different; simulations that try to 
quantify the likelihood that the local density differs from 
the average density have been done, 
e.g.,~\cite{Kamionkowski:2008vw}.
The simulations suggest the local dark matter density is within 
one to several orders of magnitude from the canonical value of 
$0.3$~GeV/cm$^3$,
with more uncertainty on the lower bound than on the upper bound,
depending on the likelihood one is willing to tolerate. 
This implies a minimum uncertainty of one to several orders of magnitude
on the direct detection cross section.  (As we will see, there
remain several additional sources of uncertainty.)

The velocity distribution is even more subtle. 
\emph{A priori} not much is known about the velocity distribution,
although there has been substantial work to estimate its form 
using $N$-body simulations \cite{Diemand:2006ik,Vogelsberger:2007ny,
Diemand:2008in,Stadel:2008pn,Springel:2008cc,Vogelsberger:2008qb}.
These simulations are generally excellent tracers of the 
\emph{average} properties of the velocity distribution
for a galaxy that is \emph{like} the Milky Way.  
They do not, however, have the level of resolution needed 
to determine the local velocity distribution.
Moreover, while most simulations do not incorporate the feedback effects 
of matter, those that do find interesting 
``dark disk'' 
structure \cite{Read:2008fh,Bruch:2008rx,Read:2009jy,Ling:2009eh}.
As has been recently emphasized in \cite{Lisanti:2010qx}, 
even isotropic spherically symmetric velocity distributions derived 
from equilibrium distributions of dark matter density result 
in departures from the Maxwellian distribution.
Local effects on the density and velocity distribution 
by solar and gravitational capture have also been considered 
in \cite{boundearth}.

Direct astronomical observations of the local stellar neighborhood
can be inferred to give information on local properties of dark matter.
The RAVE survey has constrained the local galactic escape speed
\cite{Smith:2006ym}.  The values lie in a fairly large range, 
from $460 \lsim v_{esc} \lsim 640$~km/s, depending on the parameterization 
of the distribution of stellar velocities.
Observations have also shed light on non-equilibrium local stellar motion, 
such as the possibility of ``streams'' of stars relatively nearby 
to the Sun.  Tidal streams can arise from a disrupting satellite 
dwarf galaxy or star cluster.
There has been evidence of streams in the solar neighborhood
for some time \cite{Helmi:1999uj,Chiba:2000vu,Kepley:2007vx}.
More recent work with newer stellar surveys, including RAVE, SDSS, etc.,
suggest streams are present \cite{Klement:2008ws,Klement:2009km,Smith:2009kr}
(although other work does not find evidence for local streams 
oriented along the direction perpendicular to the 
galactic disk \cite{Seabroke:2007er}).
If a tidal stream of stars provides a good tracer of 
dark matter, this or other local structure could significantly 
affect dark matter direct detection measurements
\cite{Stiff:2001dq,Freese:2003na,Savage:2006qr,Bruch:2008rx,Fairbairn:2008gz,MarchRussell:2008dy,Ling:2009cn,Kuhlen:2009vh,Lang:2010cd,McCabe:2010zh,Natarajan:2010jx,Green:2010gw}.
Generally speaking, however, prior work has considered streams
or other non-Maxwellian structure as perturbations on top of 
a (quasi-)Maxwellian distribution.

Our conclusion is that while the average density and average
velocity distribution at the Sun's galactic radius do seem
to be determined reasonably well from both observations and
$N$-body simulations, the \emph{local} dark matter density 
and \emph{local} dark matter velocity distribution remain
highly uncertain.  
Breaking with the vast majority of prior literature on dark matter 
direct detection, we consider what particle physics properties 
can be unambiguously extracted while making essentially 
no assumptions about the density and velocity distribution of 
dark matter.  

We approach this subject making the following
simplifying assumptions:
\begin{itemize}
\item[1.] One direct detection experiment reports several events 
that can be interpreted as dark matter scattering off nuclei.  
\item[2.] Nuclear recoil events are consistent with scattering off
only one type of nuclei.  The scattering process off these nuclei 
is dominated by \emph{either} a spin-independent or spin-dependent 
process, but not both.
\item[3.] No significant time variation of the rate of events is found,
and no directional detection is observed.
\end{itemize}

Since the main purpose of this paper is point out what can 
(or cannot) be extracted from a recoil spectrum independent
of local density and velocity distribution, we focus on what
can be obtained from just one experiment
(or, equivalently, treat several experiments' results
independently).
Making the second assumption is, for many experiments, 
not an assumption at all when detectors are made of
a uniform material (e.g.\ Xenon100, LUX, CDMS, etc.).
Even in experiments consisting of multiple nuclei
with very different masses (e.g.\ CRESST), 
it would require a remarkable coincidence 
to have the experiment's proportions-by-mass of different nuclei 
arranged such that dark matter were able to scatter off several of them
simultaneously with large signal over background.
For spin-dependent cross sections, typically only a few
isotopes dominate.  Again, it would be a remarkable coincidence
to have two separate scattering processes (either both spin-dependent 
or one spin-dependent and one spin-independent) contributing to
a nuclear recoil signal in a single experiment.

The third point, that we do not consider time variation 
or directional dependence in rates, is a more serious omission.  
Indeed, time variation is itself 
an intriguing signal, especially given the longstanding annual 
modulation observed by DAMA \cite{Bernabei:1998td}.  
However, we already have our work cut out for us with
the comparatively mundane time-independent signal.  Indeed, 
it is likely that convincing evidence of time variation,
or directional detection, will require years of exposure.  
Nevertheless, some work on the effect of dark matter streams
on direct detection has already been done in e.g.\
\cite{Stiff:2001dq,Savage:2006qr}.
We expect to return to this pressing issue of the purely
particle physics implications of time variation and
directional detection, independent of density or velocity distribution,
in future work.  

The upshot of these assumptions is that we consider a collection 
of nuclear recoil events and analyze their consequences.  
Given the strong bounds on spin-dependent scattering from collider
searches \cite{Beltran:2010ww}, we will mostly concentrate on
the case of spin-independent scattering.  However, our formalism
and philosophy applies to both types of scattering.

We will find that some characteristics of the recoil spectrum
are highly dependent on the astrophysical inputs, 
while others are totally independent of it.
We will, of course, extract far less information than is 
usually presented in the literature, but the information we do
extract is far more robust in the sense that it does not rely
on any properties of the local dark matter halo.  
However, since the \emph{only} way we can unambiguously probe 
the dark matter velocity distribution and abundance at the Earth's position 
is through direct detection, we argue that 
anything beyond what we claim here ultimately suffers from 
dependency on local astrophysical properties of dark matter.

\section{Kinematics}
\label{sec:kinematics}

We assume dark matter consists of a set of uncolored and uncharged 
particles $\chi_i$ with mass $m_i$, velocity distribution $f_i(\vec{v}_i)$,
and density $\rho_i$.  The dark matter velocities are taken to be
in the Earth frame.
This is a significant departure from conventional analyses.
It is much more typical to take a galactic frame velocity distribution,
motivated by simulations or simplifying assumptions about the 
galactic distribution, and boost into Earth frame.  
For us, since our velocity distributions $f(v)$ are specified 
in Earth frame, they can be used directly
for direct detection scattering.  However, since 
time-independent dark matter scattering depends only on
the \emph{magnitude} of the relative velocity, there is no way
to determine the magnitude in the galactic frame outside of
a kinematic range.  This should be kept in mind when viewing
velocity distributions later in the paper.

We will mostly consider examples with
just one dark matter candidate, although the formalism we develop
below will apply to an arbitrary set of WIMPs.
For any given dark matter particle $\chi$, it is useful to 
first consider the kinematics of the most general direct detection 
recoil event.  We consider a collision between $\chi$ with mass 
$m_\chi$, moving with velocity $\vec{v}$ in the Earth's frame, 
and a nucleus $N$ with mass $m_N$, that is stationary.  
The result of this collision, in the most general 
two-to-two case, will be an outgoing dark-sector particle, 
$\chi^\prime$, and an outgoing visible-sector particle, 
$N^\prime$.  We define $\delta_\chi \equiv m_\chi^\prime - m_\chi$
and $\delta_N \equiv m_N^\prime - m_N$, and work in the approximation
that $\delta_\chi/m_\chi,\delta_N/m_N \ll 1$, i.e., the mass differences 
between incoming and outgoing particles are small.
While generally the nucleus is not changed as a result of
WIMP scattering, there are notable exceptions, such as 
nuclear excitation \cite{Ellis:1988nb,Engel:1999kv} and rDM 
with neutron emission \cite{Pospelov:2008qx,Bai:2009cd}.

In the center-of-momentum frame, where $\vec{v}$ is the relative velocity 
of the two incoming particles, the incoming momentum 
is $\vec{p} = \mu \vec{v}$ and the outgoing momentum is 
$\vec{p}^{\, \prime}$.  Here we use the reduced mass defined 
with respect to the incoming particles,
\begin{eqnarray}
\mu \equiv \frac{m_{\chi} m_N}{m_{\chi} + m_N} \; .
\end{eqnarray}
The recoil energy of the collision is 
$E_R=q^2/2m_N^\prime$ with
\be
q^2 = p^2 + p^{\prime 2} - 2 p\, p^\prime \cos\tcom \; .
\ee  
The recoil of energy $E_R$, velocity $v$ and $\cos \theta_{\rm lab}$ 
are related by,
\bea
\frac{v^2}{2} \delta_\chi \frac{m_\chi}{m_{\chi^\prime}}
- v\frac{m_\chi}{m_{\chi^\prime}} \sqrt{2 m_{N^\prime} E_R} 
\cos \theta_{\rm lab} &&\nonumber \\
- \left[ E_R \left( 1 + \frac{m_{N^\prime}}{m_{\chi^\prime}} \right) 
+ \delta_\chi + \delta_N \right]
&=& 0~.
\eea
Define $\delta \equiv \delta_\chi + \delta_N$.  
If $\delta > 0$, we can safely perform an expansion in
$\delta/m \ll 1$ to obtain 
\be
v_{\rm min} = \frac{1}{\sqrt{2m_N E_R}} 
              \left( \frac{m_N E_R}{\mu} + \delta \right)~.
\label{eq:vmin}
\ee
which taking $\delta_N \ra 0$ is the well-known result for inelastic 
dark matter (iDM) \cite{TuckerSmith:2001hy,TuckerSmith:2004jv,Chang:2008gd}.
By ``safe'' we mean that our upper bound
on $v_{\rm min}$, which is in the far non-relativistic regime, 
automatically implies $|\delta| \ll m_\chi,m_N$
to allow scattering to be kinematically possible.

Up to higher order terms in $\delta/m$, we obtain an expression for 
the recoil energy 
\be
E_R^2 + 2 E_R \frac{\mu}{m_N} (\delta - \mu v^2 \cos^2\tlab )
+ \frac{\mu^2}{m_N^2} \delta^2 = 0
\label{eq:ER}
\ee
The recoil energy is unique for a given fixed scattering
relative velocity $v$ and nucleus recoil angle $\tlab$
and can be solved by the usual quadratic formula,
\bea
E_R &=& \frac{\mu}{m_N} \left[
        \left( \mu v^2 \cos^2\tlab - \delta \right) \right. \\
       &&{}\left. \pm (\mu v^2 \cos^2\tlab)^{1/2} 
        \left( \mu v^2 \cos^2\tlab - 2 \delta \right)^{1/2} \right] \nonumber
        \;  .
\label{eq:ERsoln}
\eea
This result has the well known feature that the
smallest recoil energies come from maximizing $v^2 \cos^2\tlab$, 
corresponding physically to head-on collisions at the 
highest velocities available.

\section{Event Distributions}

Our basic assumptions consist of assuming the scattering process 
is off only one type of nuclei.
We will, however, remain general with respect to the possibility
of multiple WIMPs with different masses, abundances, and cross sections.  
One might think it requires a large coincidence to
have several dark matter particles with cross sections
large enough to produce events in an experiment.
However, there are well known counterexamples where it can be
natural to have the abundance of particles to be independent of
their mass (and thus, have several candidates of different
masses with similar abundances, using for example the WIMPless miracle
\cite{Feng:2008ya}).  

The event rate of dark matter scattering \cite{Lewin:1995rx}, 
differential in $E_R$, is determined by 
\be
\label{eq:rateeq}
\frac{d R}{d E_R} = \sum_i \frac{N_T \rho_{\chi_i}}{m_{\chi_i}} 
\int_{v_{i,\mathrm{min}}}^{v_{\mathrm{max}}}
d^3 \vec{v}_i\, f_i(\vec{v}_i(t)) 
\frac{d\,\sigma_{i} |\vec{v}_i|}{d E_R} ~,
\ee
where the sum is over different species of WIMPs, 
$m_N \simeq A m_p$ is the nucleus mass with $m_p$ the proton
mass and $A$ the atomic number.  The recoil energy depends on the 
kinematics of the collision, as described above.  
Given our assumption of no significant time variation in the rate,
$f(\vec{v}_i(t)) \ra f(\vec{v}_i)$, and thus 
we are effectively neglecting the Earth's motion around the Sun.
This is a reasonable approximation so long we are probing  
velocities larger than Earth's velocity in the Sun's frame,
i.e., $v_{\rm max} \gsim 30$~km/s.  
Typically the maximum speed is taken to be 
$v_{\rm max} = v_{\rm earth} + v_{\rm esc}$, the galactic escape 
velocity boosted into the Earth frame.  However, $v_{\rm max}$ is 
ultimately determined by the (unknown) details of the dark matter 
velocity distribution in Earth frame.  

Given our assumption of no direction dependent signal, 
we can carry out the angular integral in 
Eq.~(\ref{eq:rateeq}), reducing it to a one dimensional integral
where we introduce the 
quantity\footnote{The velocity distribution is normalized 
such that $\int d^3 v f(v) = 1$.} $f_1(v)=\int d\Omega f(\vec{v})$.  
The differential rate becomes
\bea
\label{eq:oneDrate}
\frac{d R}{d E_R} &=& \sum_i \frac{N_T \rho_{\chi_i} m_N}{\mu_i^2 m_{\chi_i}} 
F_N^2(E_R) \nonumber \\
&&{}\times 
   \int_{v_{i,\mathrm{min}}}^{v_{\mathrm{max}}} d v_i \, v_i f_{i1}(v_i) 
   \bar\sigma_{i}(v_i,E_R)~,
\eea
where we have written
\bea
\frac{d \sigma_i}{d E_R} &=& F_N^2(E_R) \frac{m_N}{\mu_i v_i^2} \bar\sigma_{i}
(v_i,E_R)
\eea
in terms of the nuclear form factor $F^2_N (E_R)$.
There are several possible forms for the scattering cross section 
$\bar\sigma_{i}(v,E_R)$, depending on the interaction,
\be
\bar\sigma_{i}(v,E_R) = \left\{ \begin{array}{l} 
                                \sigma_{i0} \\
                                \sigma_{i0} F_{\chi_i}^2(E_R) \\
                                \sigma_{i0}(v) F_{\chi_i}^2(E_R) \\
                                \sigma_{i0}(v,E_R)
                                \end{array} \right. \; .
\label{eq:factorize}
\ee
The different forms for $\bar\sigma$ correspond to functional
forms of known dark matter scattering that contain velocity
and/or recoil energy dependence.  The first possibility,
a constant independent of $v$ and $E_R$ is the well-known
isotropic ($s$-wave) cross section that results at lowest order in the
non-relativistic expansion from many
dark matter models.  

The second possibility contains a dark matter form factor $F_{\chi_i}$
(following the standard normalization convention 
$F_{\chi_i}(E_R = 0) = 1$) and commonly occurs in models
of composite dark matter \cite{Nussinov:1985xr,Alves:2009nf,Kribs:2009fy}. 
Our formalism will handle the factorizable forms,
i.e., the first three of Eq.~(\ref{eq:factorize}), which incorporates
the vast bulk of what has been considered in the literature.
We will not, however, consider the cross sections that contain 
completely arbitrary nonfactorizable velocity and recoil energy dependence
[c.f., the most general form written on the fourth line
of Eq.~(\ref{eq:factorize})].

We now turn to the question of what can be inferred from a signal in 
direct detection experiments using (\ref{eq:oneDrate}) without making 
any assumptions about $f_1$ or the dark matter scattering cross section 
$\sigma_0$.  We will however, make an assumption about the 
maximum dark matter speed, $v_{\rm max}$, and we will demonstrate how the 
derived dark matter properties depend on this assumption.

\section{Deconvoluted Scattering Rate}

Since the scattering rate (\ref{eq:oneDrate}) in any given
direct detection experiment is proportional to the
nuclear form factor, we first factor it out.  This leads to a 
definition of a new quantity, $\mathcal{R}$, that we call the 
``deconvoluted scattering rate'' -- deconvoluted of the nuclear 
form factor,
\bea
\!\!\!\!\!\!\!\!\! \mathcal{R} &\equiv &
\frac{1}{F_N^2(E_R)} \frac{d R}{d E_R} \nonumber \\
&=& \sum_i {\cal N}_i m_N 
\int_{v_{i,\mathrm{min}}}^{v_{\mathrm{max}}} d v_i \, v_i f_{i1}(v_i) 
\bar\sigma_{i}(v_i,E_R).
\label{eq:deconvolutedeq}
\eea
Some overall factors have been buried into a normalization factor,
$\mathcal{N}_i = N_T \rho_{\chi_i}/(\mu_i^2 m_{\chi_i})$.
While there are important uncertainties in the determination
of dark matter nuclear form factors from nuclear data \cite{Ellis:2008hf},
this is not our concern.  Errors on
the deconvoluted scattering rate ought to take into account nuclear 
form factor uncertainties.  

Next, taking a derivative with respect to $E_R$ we find
\bea
\frac{d \mathcal{R}}{d E_R} 
&=&  \sum_i {\cal N}_i m_N 
\left(\int_{v_{i,\mathrm{min}}}^{v_{\mathrm{max}}} d v_i v_i f_{i1}(v_i) 
\frac{d \bar\sigma_{i}(v_i,E_R)}{d E_R}  \right.\nonumber \\
&&\left.
\!\!\!\!\!\!\! -v_{i,{\rm min}} \frac{d v_{i,{\rm min}}}{d E_R}
f_{i1}(v_{i,{\rm min}}) \bar{\sigma}(v_{i,{\rm min}},E_R)
\right).
\label{eq:fgeneral}
\eea
For arbitrary $2 \ra 2$ kinematics (elastic or inelastic), 
we can replace
\bea
v_{i,{\rm min}} \frac{d v_{i,{\rm min}}}{d E_R} &=&
 \frac{m_N^2 E_R^2 - \mu_i^2 \delta_i^2}{4m_N \mu_i^2 E_R^2} \; .
\eea
This is as far as we can go with a general signal from an
ensemble of WIMPs with arbitrary cross sections.  

For a single WIMP with a factorizable cross section, 
Eq.~(\ref{eq:deconvolutedeq}) can be used to solve for 
$f_1(v)$ (see also \cite{Drees:2007hr,Drees:2008bv,Chou:2010qt,Shan:2010qv}):
\begin{widetext}
\be
f_1(v_{\rm min}(E_R)) =
    - \frac{4 \mu^2 E_R^2}{m_N^2 E_R^2 -\mu^2 \delta^2} 
    \frac{1}{\mathcal{N} \sigma_0(v_{\rm min}(E_R)) F_\chi^2(E_R)} 
    \left(
    \frac{d \mathcal{R}}{d E_R}  -
    \mathcal{R}
    \frac{1}{F_\chi^2(E_R)}
    \frac{d F^2_\chi(E_R)}{d E_R} 
    \right)~.
\label{eq:ffrominversion}
\ee
\end{widetext}
This result allows us to gain information on the 
velocity distribution of dark matter evaluated at the 
\emph{minimum} velocity to scatter for a given 
recoil energy $E_R$.  With scattering data over the 
range $E_R^{\rm min} < E_R < E_R^{\rm max}$, 
we obtain information on the velocity 
distribution $f(v)$ over a range of $v$:  
$v_{\min}(E_R^{\rm min}) < v < v_{\rm min}(E_R^{\rm max})$.

For an ensemble of WIMPs, $\chi_i$, without dark matter form factors, 
the inversion result can be written as
\be
\frac{d \mathcal{R}}{d E_R} = \sum_i w_i(v,E_R) f_{i1}(v) ~, 
\label{eq:fensemble}
\ee
where the velocity distributions of the WIMPs are ``weighted'' by
the factors 
\be
w_i(v,E_R) = 
    - \frac{1}{4}
    \left( \frac{m_N^2}{\mu_i^2} 
           - \frac{\delta_i^2}{E_R^2} \right)
    \mathcal{N}_i
    \sigma_{i0}(v) 
\ee
For an ensemble of WIMPs \emph{with} form factors, no simple
closed form can be written.

\section{$f$-condition}

There is valuable information that can be extracted from
Eqs.~(\ref{eq:ffrominversion}) and (\ref{eq:fensemble}).  
We know the velocity distribution
of dark matter must be positive for all $v$, 
\be
f(v) \ge 0~,
\label{eq:f-condition}
\ee
which we call the ``$f$-condition".  Using this condition, 
the right-hand side of Eq.~(\ref{eq:ffrominversion}) must be positive.  
Similarly the $f$-condition also places constraints on the terms appearing 
in Eq.~(\ref{eq:fensemble}).

Consider the case of single WIMP with standard elastic scattering 
without a dark matter form factor, $\delta = 0$ and $F_\chi^2(E_R) = 1$.  
From Eq.~(\ref{eq:ffrominversion}) we conclude that the 
deconvoluted scattering rate is always a decreasing function of $E_R$.  

A more striking consequence is reached if a rising 
deconvoluted scattering rate is ever observed.
Should there be a range of data 
where the deconvoluted scattering rate satisfies
$d\mathcal{R}/d E_R > 0$,
this would signal the presence of either an inelastic threshold for 
scattering or a dark matter form factor (or both).  
This can be seen most easily by classifying which terms 
in Eq.~(\ref{eq:ffrominversion}) can be positive or negative.

This general characteristic of the shape of the nuclear recoil 
distribution can thus be used to provide an experimental signal 
of non-standard dark matter interactions, \emph{independent} of the 
local velocity distribution and local density of dark matter.
This observation is one of main conclusions of our paper.

Eq.~(\ref{eq:ffrominversion}) illustrates how the introduction of 
non-trivial particle physics can allow for structure in the 
recoil spectrum while still satisfying Eq.~(\ref{eq:f-condition}),
through either an inelastic cross section or one with a
non-trivial dark matter form factor.
One smoking gun for inelastic dark matter is a rise and fall 
of the (deconvoluted) dark matter scattering rate.
Yet, Eq.~(\ref{eq:ffrominversion}) makes clear that a dark matter 
form factor whose functional form has positive powers of $E_R$ 
over a range of $E_R$ within the experimental sensitivity 
also can lead to a rise and fall of the deconvoluted scattering rate. 
Since no direct detection experiment can probe $F_\chi^2(E_R = 0)$, 
the functional form of a dark matter form factor 
is essentially unconstrained.\footnote{We are not aware
of a way to obtain $F_\chi^2(E_R) \propto E_R^n$ for $n < -2$ 
from a microscopic theory.
However, since this does not lead to $d\mathcal{R}/dE_R>0$, this 
limitation is not of much concern.}
Indeed, Ref.~\cite{Feldstein:2009tr,Chang:2009yt} demonstrated that specific
models of what they called ``form factor dark matter'' can 
reproduce a rising deconvoluted scattering rate, $d\mathcal{R}/d E_R > 0$, 
like inelastic dark matter.  
It is interesting to consider simple functional forms for the 
form factor $F^2_\chi(E_R)$, and whether they can (or cannot) fake 
inelastic dark matter given an arbitrary velocity distribution.  
We carry out this study in Sec.~\ref{sec:ffidm}.

The $f$-condition also allows us to see that 
there is a complete degeneracy between $\sigma(v)$ and $f_1(v)$.
No one experiment can separate a velocity-dependent
dark matter scattering cross section from astrophysical 
structure in the velocity distribution.  To the extent that 
multiple experiments do not have the same $\sigma(v)$ 
for different nuclei, it should be possible to break the degeneracy 
through multiple observations.

\begin{figure*}[t] 
   \centering
   \hspace*{0.05\textwidth}
   \includegraphics[width=0.40\textwidth]{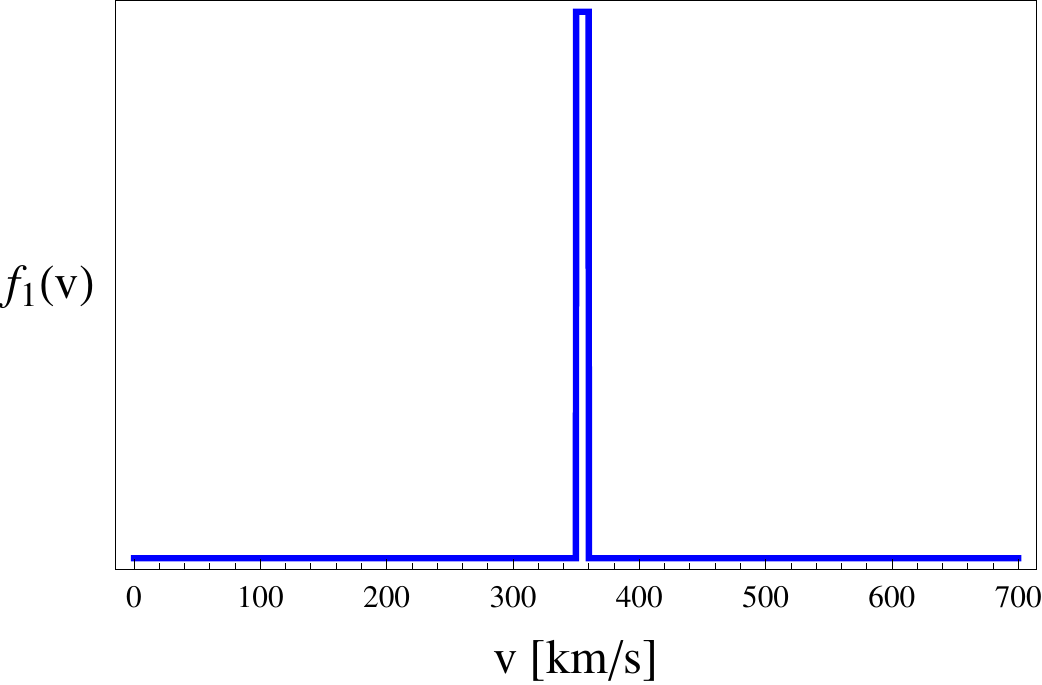}    \hfill
   \includegraphics[width=0.40\textwidth]{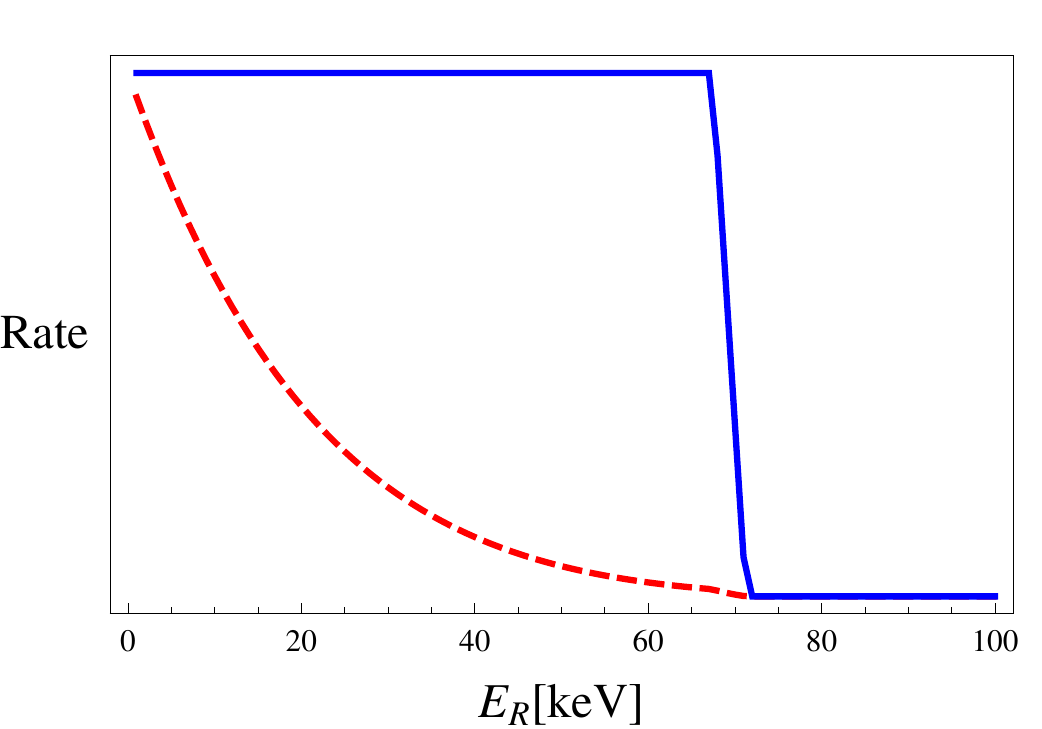}
   \hspace*{0.05\textwidth} \\
   \hspace*{0.05\textwidth}
   \includegraphics[width=0.40\textwidth]{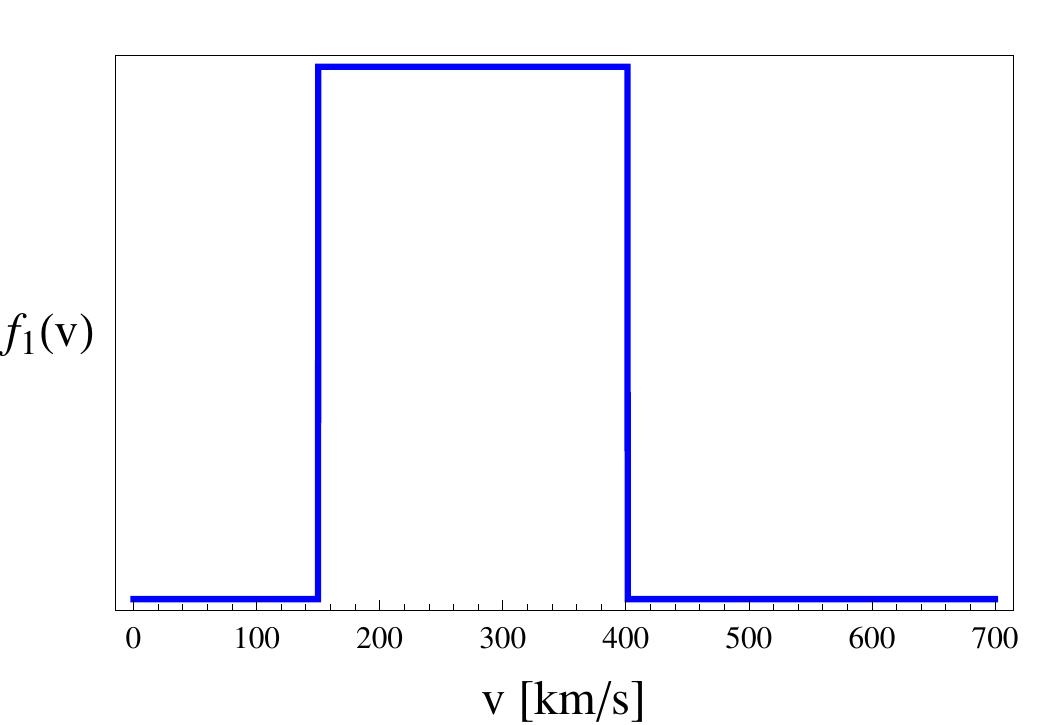}     \hfill
   \includegraphics[width=0.40\textwidth]{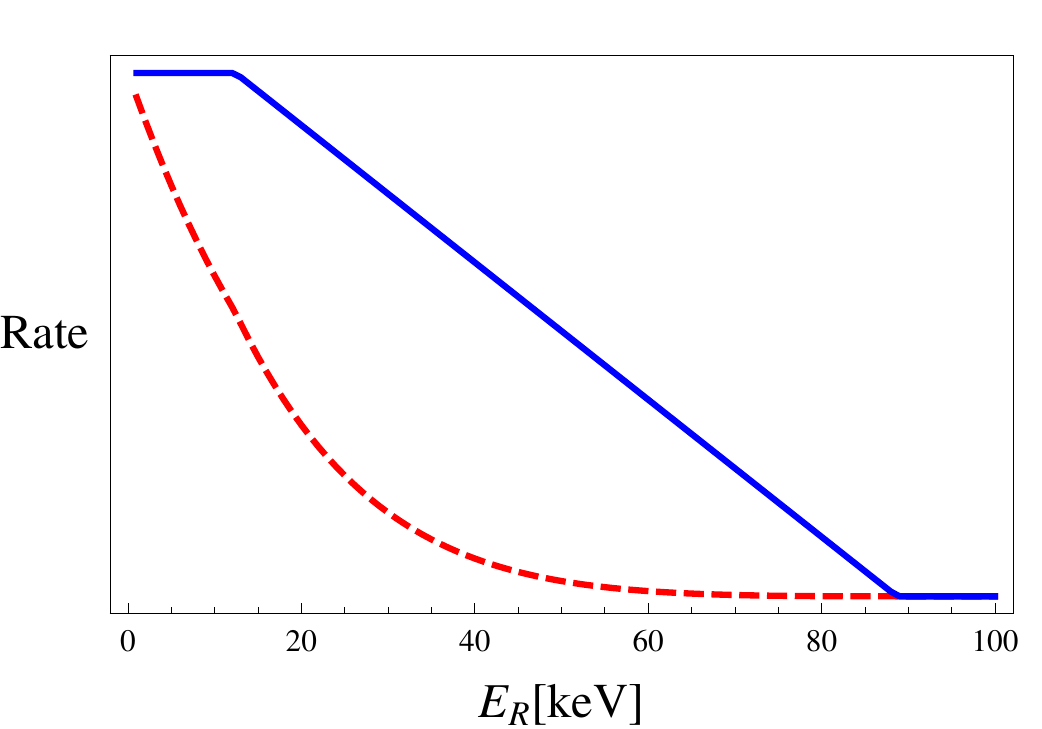}  
   \hspace*{0.05\textwidth} \\
   \hspace*{0.05\textwidth}
   \includegraphics[width=0.40\textwidth]{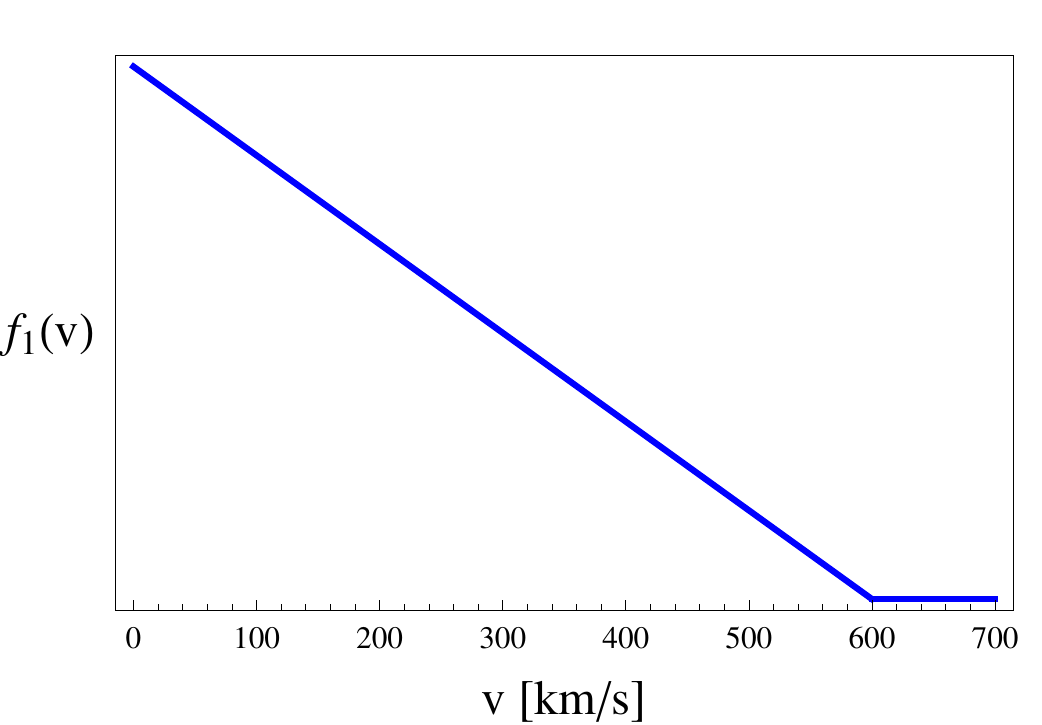}    \hfill
   \includegraphics[width=0.40\textwidth]{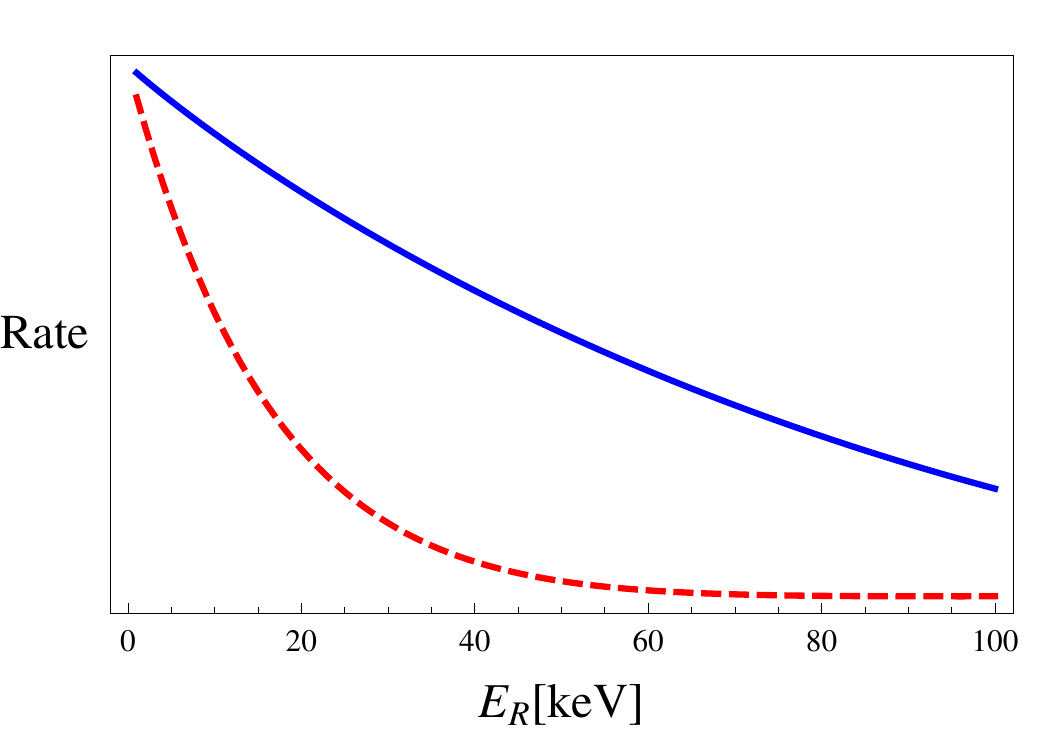} 
   \hspace*{0.05\textwidth} 
   \caption{Some examples of the relationship between velocity distribution 
(LH plots) and observed recoil spectrum, $d R/d E_R$, (red dashed in RH plots) 
and deconvoluted spectrum, $\mathcal{R}$, (blue solid in RH plots).  
The dark matter mass is taken to be 100 GeV with elastic scattering off 
xenon.}
   \label{fig:exampleftorate}
\end{figure*}

It is instructive to consider several possible forms of $f_1(v)$ 
and how they map into distributions of energies.  
As a simple starting point, consider a velocity distribution 
that can be approximated by a delta function,
\bea
f_1(v) = \delta(v - v_0)~.
\eea
This could correspond physically to  
a local density of dark matter that
is contained within a stream moving with speed $v_0$
in the Earth's frame, or a scattering cross section with a 
resonant feature~\cite{Bai:2009cd}.  
From Eq.~(\ref{eq:deconvolutedeq}), the deconvoluted spectrum 
will be a constant provided $v_{min} \leq v_0$ and zero otherwise.  
In terms of $E_R$, the rate is constant provided $E_R$ 
is below a certain threshold determined by $v_0$, and zero above it.  
This result is shown graphically for elastic scattering off xenon 
in the top panels of Figure~\ref{fig:exampleftorate}.

One can build a discretized version of any $f_1(v)$ by combining a 
suitable number of delta functions.  Every delta function contributes 
to the energies below its threshold $E_R$, and zero above it.  
In the case of elastic scattering, the result is a deconvoluted distribution
which falls off at larger energies.  Two other simple cases, 
$f_1(v) = $~constant for some range of $v$ and $f_1(v)$ 
falling linearly in $v$ are also shown in
Fig.~\ref{fig:exampleftorate}.

Several further comments are in order.  If we imagine a spectrum
of several delta functions, the lowest velocity contributions may 
result in recoil energies \emph{below} an experiment's detection 
threshold, and therefore obviously cannot be observed.  
Given that the weights of these delta functions are also undetermined, 
this means that the normalization of the velocity distribution 
cannot be experimentally determined.  

Experiments may also be limited in sensitivity at for some recoil
energy $E_R > E_R^{\rm exp,max}$.  Contributions to the velocity
distribution at very high velocities that contribute to the
recoil spectrum in this range would also not be fully probed
by an experiment.  This is another source of uncertainty 
regarding the normalization of the velocity distribution.

The upshot of this is that there is both \emph{astrophysics} 
as well as fundamental \emph{experimental sensitivity} limitations
in interpreting the overall normalization of a given recoil
spectrum as arising from a given cross section.  This limitation
is fundamental -- for an arbitrarily large fraction of dark matter 
moving slowly enough in Earth frame, no experiment will be sensitive
to such nuclear recoils, and thus, no experiment can measure
a cross section independently of astrophysics.  
However the results we find here are based on the shape
of the recoil spectrum and not the normalization, and thus
can be determined independent of astrophysics.

\section{Case Studies}

We now consider several examples of what can, and cannot, be
determined from a nuclear recoil spectrum of a single 
dark matter detection experiment.  Given the discussion above,
in all cases the normalization of $f_1(v)$ (and thus the cross section)
is completely arbitrary.

\subsection{Elastic Dark Matter}

\begin{figure}
\centering
\includegraphics[width=0.49\textwidth]{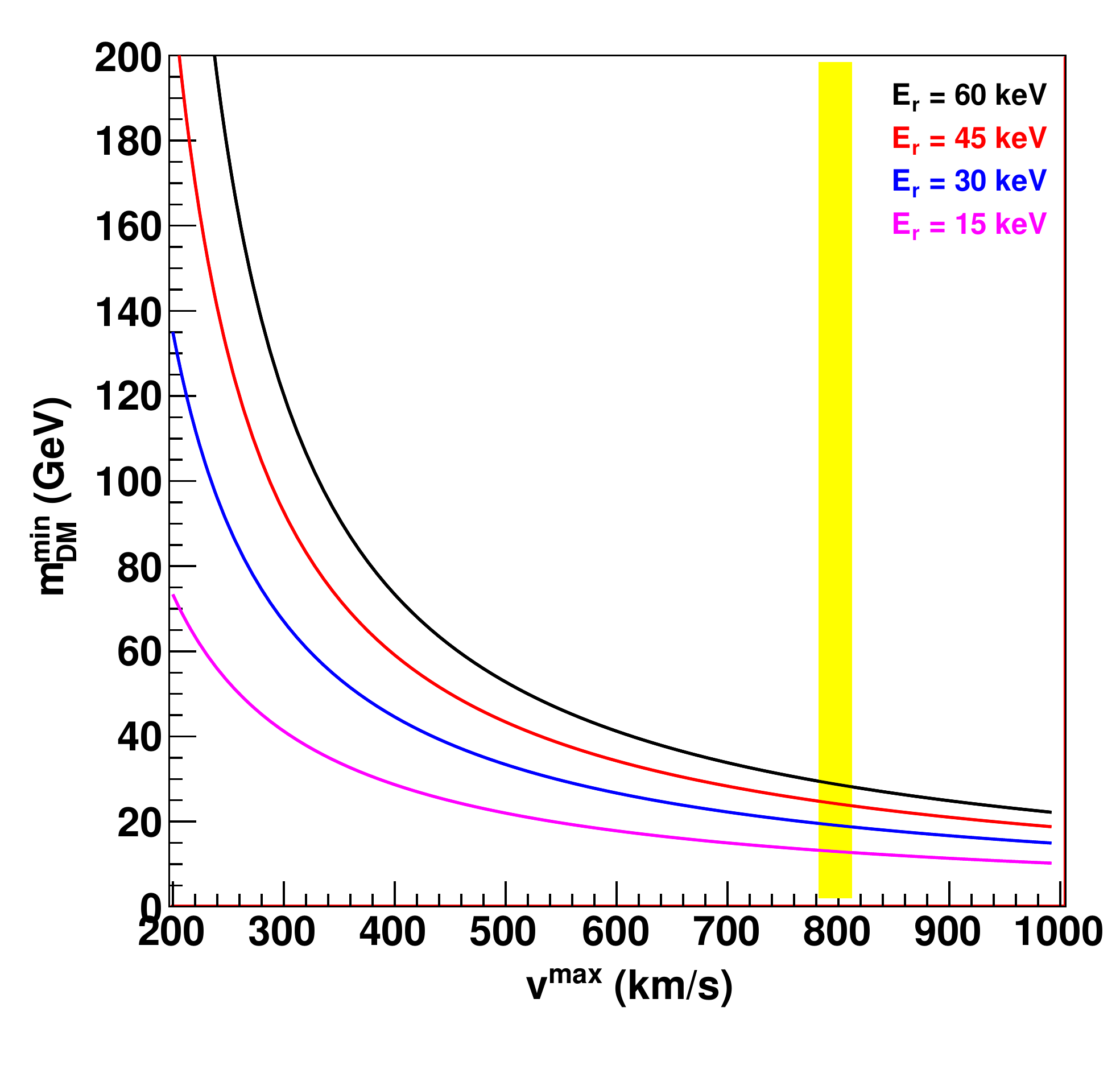}
\caption{The kinematic bound on the dark matter mass, assuming 
elastic scattering off xenon.  Different curves are plotted for
different assumptions about what events are assumed to
be the highest energy nuclear recoils from dark matter.  The yellow band
corresponds to the typically largest relative velocity
arising from summing 
$v_{\rm esc} + v_{\rm earth} \lsim 800$~km/s.}
\label{fig:kinematicsxenon10}
\end{figure}

For the case of elastic dark matter (eDM), defined as 
standard elastic scattering $( \delta =0 )$ with no dark matter 
form factor $( F^2_\chi(E_R) = 1)$, we can determine
a lower bound on $m_{\chi}$ provided one has knowledge of 
the maximum possible velocity, $v_{\rm max}$. 
From Eq.~(\ref{eq:ER}), we can solve for $\mu$ as a function
of $E_R$ and the other parameters.
For elastic scattering, the result is particularly simple, 
\bea
\mu_{\rm min} &=& \sqrt{\frac{m_N E_R}{2 v_{\rm max}^2}} \; , 
\eea
demonstrating that the strongest lower bound on the dark matter mass 
comes from the highest recoil energy events at the maximum
dark matter velocity (in Earth frame).
We illustrate this bound in Fig.~\ref{fig:kinematicsxenon10}, 
by showing the bound on $m^{\rm min}_{\chi}$ as a function of 
$v_{\rm max}$ for four possible values of the maximum recoil energy 
from a distribution of events where dark matter scatters off xenon.
In the next section, we will see that the analogous constraints
on $m^{\rm min}_{\chi}$ for inelastic dark matter depends
on the inelastic threshold.

The deconvoluted scattering rate, Eq.~(\ref{eq:ffrominversion}), 
takes on the simple form in eDM,
\be
\label{eq:ffrominversionEDM}
f_1(v) = -\frac{4E_R}{m_N^2\mathcal{N}\tilde{\sigma}_0(v)} 
         \frac{d\mathcal{R}}{dE_R}~.
\ee
The positivity of $f(v)$ (and $\tilde{\sigma}_0(v)$) means that for 
elastic scattering the spectrum of recoil events \emph{must} be a 
monotonically decreasing function of energy.  If this is not observed, 
we can immediately rule out elastic scattering, completely 
independently of any assumptions about how it couples to nuclei or 
how it is distributed in our galaxy.  

We now discuss what can be determined if indeed a falling spectrum 
is observed.  As a surrogate for experimental data, and to demonstrate 
our technique, we generate pseudo-data for a 100 GeV WIMP 
elastically scattering off xenon, 
assuming the standard Maxwellian distribution for the 
dark matter velocity in galactic frame.    
Specifically, our input contains a distribution with 
characteristic speed, $v_0=220$ km/s (in galactic frame), with an 
escape speed of $v_{esc}=500$ km/s (in galactic frame).
Since we do not consider time-dependent signals, we take the 
Earth's velocity to be a constant, $v_{earth}=230$ km/s (in galactic frame).  
We consider the lower threshold of our pseudo-experiment to be 
recoil energies of 5 keV and the upper threshold we take to be 80 keV
(the latter happens to be below the first zero of the nuclear 
form factor of xenon).  These specifications are a reasonable 
approximation of the capabilities of the 
Xenon100 experiment~\cite{Aprile:2010um}.  We ignore detector effects 
such as energy resolution and efficiency and assume that these 
are sufficiently well known that the recoil energy distribution of
Eq.~(\ref{eq:rateeq}) can be determined from the data.  

\begin{figure}[t]
\begin{center}
\includegraphics[width=0.49\textwidth]{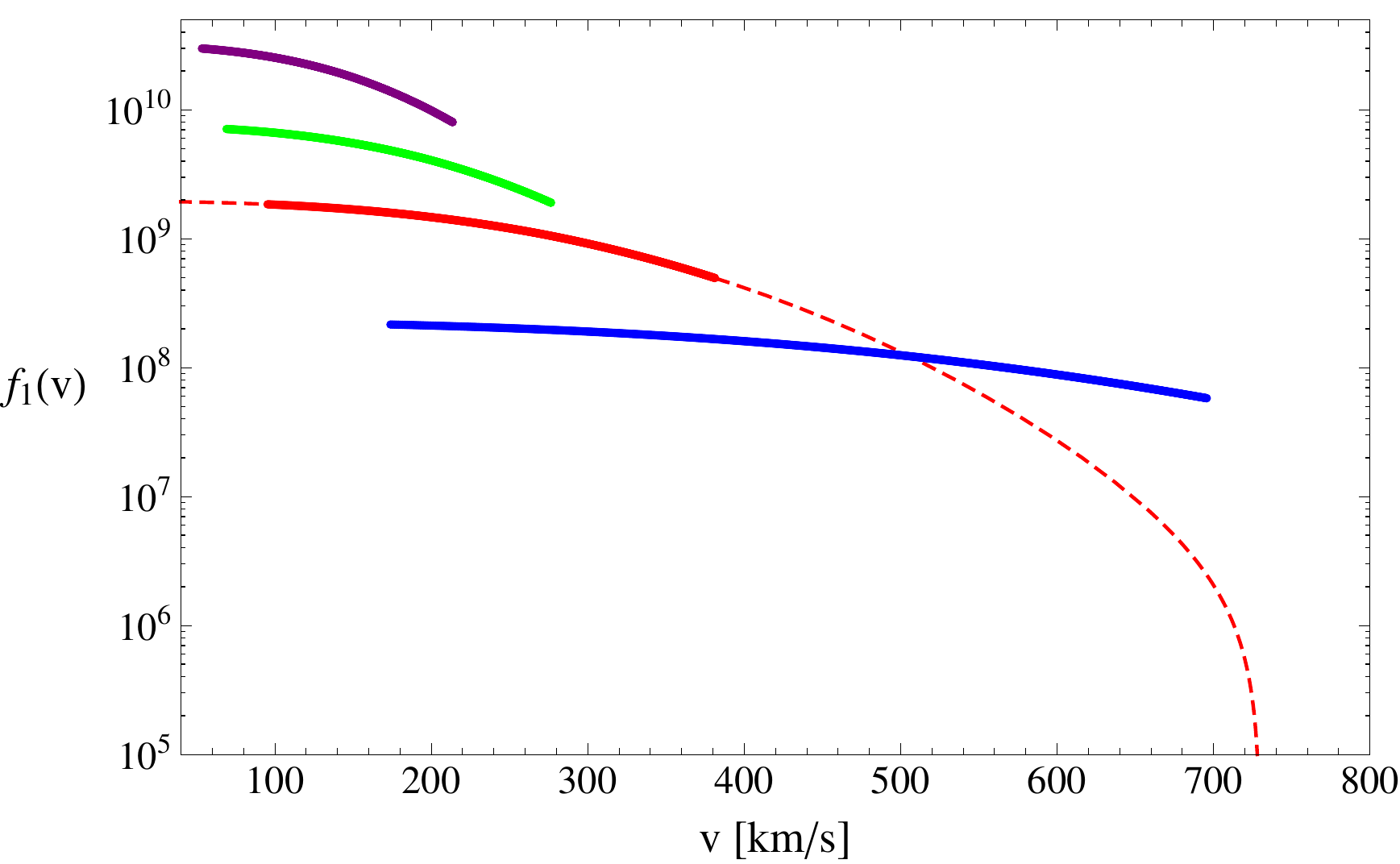}
\caption{Velocity distributions derived from pseudo-data, 
with $5\ \kev \le E_R \le 80\ \kev$, generated for a 100 GeV WIMP 
elastically scattering of xenon.  The data is generated assuming a 
Maxwellian distribution for the dark matter, dashed red line.  
For $m_\chi=40, 100, 500$~GeV (thick line segments, from bottom to top) 
the derived velocity distributions, $f_1(v)$, are shown.}
\label{fig:eDM}
\end{center}
\end{figure}

Next, we invert this ``data'' using Eq.~(\ref{eq:ffrominversionEDM}) 
and assume the dark matter is scattering elastically.  
For simplicity, we use the same scattering cross section used to 
generate the data.  This only affects the normalization of $f_1(v)$ 
and does not alter any of our arguments.  In Fig.~\ref{fig:eDM} 
we show the derived velocity distributions $f_1(v)$ for various choices 
of dark matter mass.  As expected, for the correct choice of 
dark matter mass the derived velocity distribution agrees with 
that used to generate the data.  However, since data is only taken 
over a finite range of recoil energies the velocity distribution 
is only known over a finite range of velocities, corresponding to 
the $v_{min}$ associated with the $E_R$, Eq.~(\ref{eq:vmin}).

As was discussed in Sec.~\ref{sec:kinematics} it is possible to 
place a lower bound on the dark matter mass by assuming a maximum speed 
for dark matter in our halo, the highest energy recoil events, 
taken to be 80 keV in the above, then determine a minimum mass 
through Eq.~(\ref{eq:vmin}).  However, as can be seen from 
Fig.~\ref{fig:eDM} no such upper bound can be made.  
For all assumed masses used in Fig.~\ref{fig:eDM} the resulting 
velocity distributions appear \emph{a priori} to be perfectly reasonable: 
$f_1(v)$ is positive and finite.  Without further model-dependent 
assumptions, or additional experimental results 
(either from another experiment or from raising the upper threshold 
at the first experiment), all dark matter masses, 
$m_\chi\ge m^{min}_\chi$, give a reasonable fit to data.  
Using Eqs.~(\ref{eq:vmin}) and (\ref{eq:ffrominversionEDM})
in the limit of very heavy dark matter, the derived velocity 
distribution becomes independent of the dark matter mass and 
only depends on the target.  

\subsection{Elastic Dark Matter: CoGeNT}

A further interesting example of elastic dark matter is the
recent observation of an excess of low energy recoil events 
by CoGeNT \cite{Aalseth:2010vx}.
The CoGeNT collaboration demonstrated that this is 
consistent with light dark matter,
$7$-$10$ GeV, where the mass was determined by taking
the standard Maxwellian velocity distribution.

\begin{figure}[t]
\begin{center}
\includegraphics[width=0.49\textwidth]{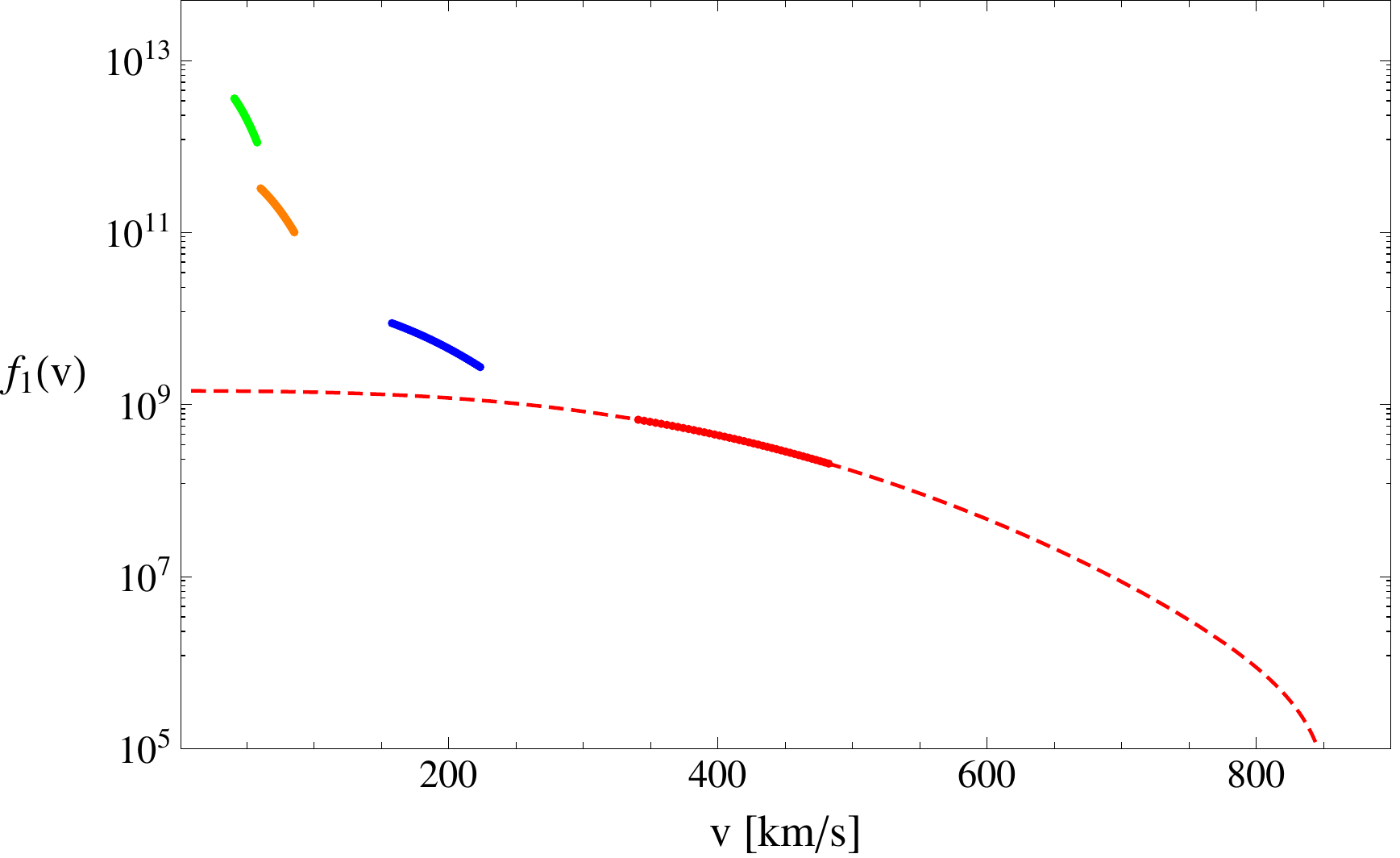}
\caption{Velocity distributions derived from the CoGeNT data.
The dashed red line is the ordinary Maxwellian velocity distribution
for an elastically scattering dark matter particle that fits the 
CoGeNT excess with a mass of $8$~GeV\@.
The derived velocity distributions, $f_1(v)$, are shown
for $m_\chi=8, 20, 100, 500$~GeV 
(thick line segments, from bottom to top).}
\label{fig:cogent}
\end{center}
\end{figure}

Following our procedure above, we have taken a fit to the 
low recoil energy excess at CoGeNT \cite{Hooper:2010uy}
and reverse-engineered 
the distribution to determine the needed velocity distribution
for (much) larger dark matter masses in Fig.~\ref{fig:cogent}.
It is easy to understand the shift in the velocity
distributions.  For light dark matter, $m_\chi<m_{\rm Ge}$, 
the maximum recoil energy 
is $E_R^{\rm max} \simeq m_{\chi}^2 v_{\rm max}^2/m_{\rm Ge}$.  
Once the dark matter
mass is larger than about $m_{\rm Ge}$, the maximum recoil
energy asymptotes to $E_R^{\rm max} \simeq m_{\rm Ge} v_{\rm max}^2$.
The shift in $v_{\rm max}$ from $m_\chi \simeq 7$~GeV to 
$m_\chi \gg m_{\rm Ge}$ is thus a factor of $\simeq 10$.
Indeed, from Fig.~\ref{fig:cogent} we see that the largest 
minimum velocity shifts from about $500$~km/s to about $60$~km/s
as the mass is increased from $8$~GeV to $500$~GeV\@.

By itself, the CoGeNT data can thus be fit by \emph{any} dark matter
candidate above the kinematic minimum, which for $v_{max} \sim 800$ km/s 
is about $6$ GeV\@.
The conclusion that CoGeNT implies ``light dark matter'' 
is thus not warranted if the local dark matter density and 
velocity distributions can be freely adjusted.

The obvious objection to the large dark matter, small velocity
distribution dark matter interpretation of the CoGeNT data
is that other direct detection experiments should be 
sensitive to larger masses, and potentially rule it out.  
We can test this assertion by translating the CoGeNT observed
energies into nuclear recoil energies (following \cite{Essig:2010ye}),
and then compute the predicted spectrum at an experiment
using xenon.  This assumes, of course, that whatever 
scattering process is occurring in germanium also occurs
in xenon.  

\begin{figure}[t]
\begin{center}
\includegraphics[width=0.49\textwidth]{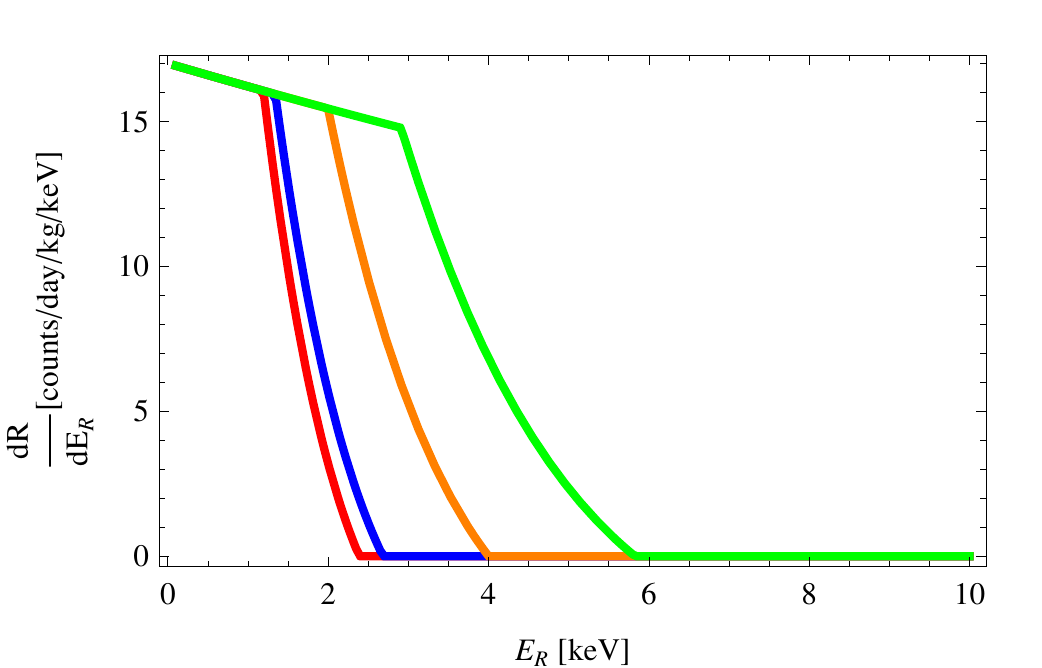}
\caption{Predicted recoil spectra at a Xenon experiment for the
velocity distributions that fit the CoGeNT excess shown in 
Fig.~\ref{fig:cogent}.  The lines correspond to $m_\chi = 8,20,100,500$
from left to right.  Notice that the normalization corresponds 
to thousands of events at Xenon10 or Xenon100.}
\label{fig:cogent2xenon}
\end{center}
\end{figure}

In Fig.~\ref{fig:cogent2xenon} we show the predicted recoil
spectrum at a xenon experiment, for each of the velocity 
distributions show in Fig.~\ref{fig:cogent}.
In each case, the prediction is large numbers of recoil
events below $6$~keVnr.  This may or may not be allowed
by existing data from Xenon10 \cite{Angle:2007uj}
and Xenon100 \cite{Aprile:2010um}.
The principle difficulty is determining the sensitivity
of these experiments to very low recoil energies,
specifically the conversion factor $L_{\rm eff}(E_R)$.  
There has been a spirited discussion on this point
\cite{Collar:2010gg,Collaboration:2010er,Collar:2010gd,Savage:2010tg,Collar:2010nx,Sorensen:2010hq,Collar:2010ht}.
Clearly, masses much heavier than considered by
CoGeNT are allowed, while arbitrarily large masses
(represented by $500$~GeV) may be constrained by this 
existing or future xenon data.  In any case, our central 
conclusion is that qualitatively larger dark matter masses 
can fit the CoGeNT data if the velocity distribution is adjusted
as we illustrated above.

\subsection{Inelastic Dark Matter}

Due to the splitting of inelastic dark matter, $v_{min}$ is not 
monotonic in $E_R$, which means that the highest energy recoil events 
may not give the strongest bound on the dark matter parameters.  
The expression for $v_{min}$, Eq.~(\ref{eq:vmin}), has a minimum, 
and correspondingly the deconvoluted recoil spectrum, $\mathcal{R}$, 
has a peak at 
\be
E_R^{\rm peak} = \frac{m_\chi}{m_\chi+m_N}\delta~,
\label{eq:recoilpeak}
\ee
the observed spectrum, $d R/d E_R$, 
typically has a peak at a lower energy due to 
the nuclear form factor.  Whether the highest or lowest energy bins 
of an experiment place the strongest constraints on $m_\chi$ and 
$\delta$ depends on where this peak falls.  

\begin{figure*}[t]
\begin{center}
\hspace*{0.05\textwidth}
\includegraphics[width=0.40\textwidth]{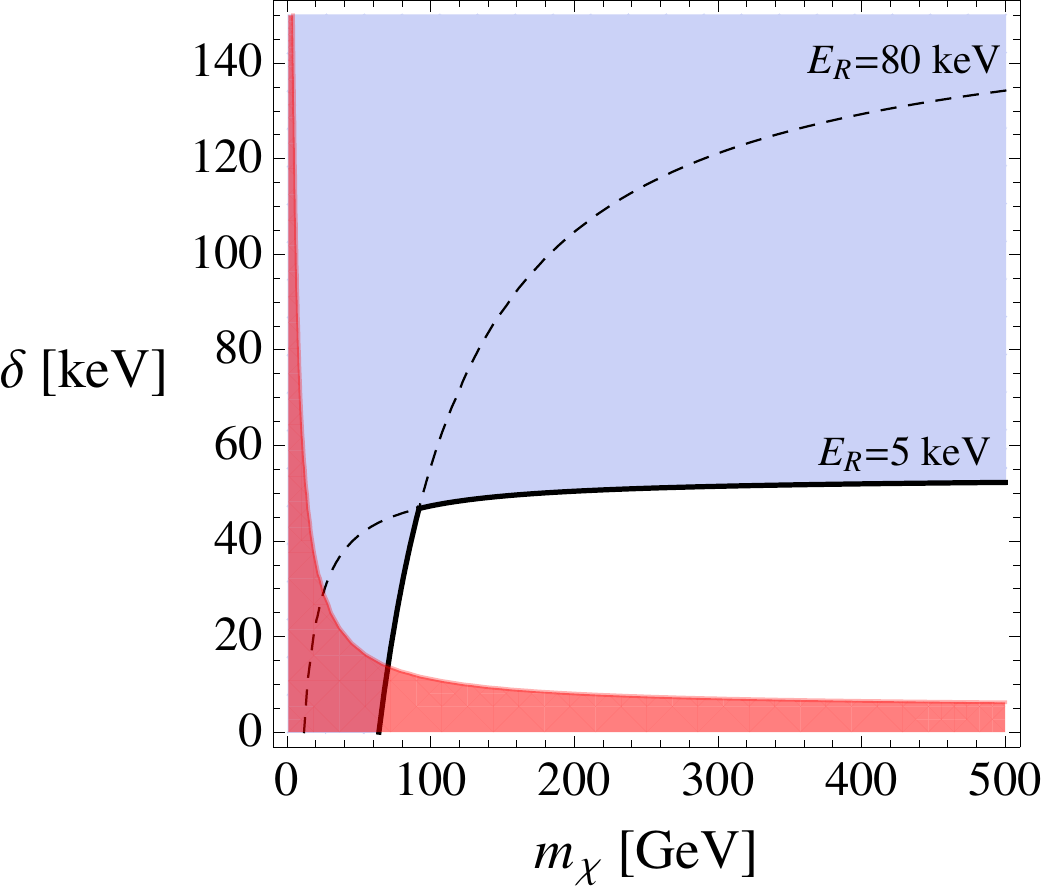} \hfill
\includegraphics[width=0.40\textwidth]{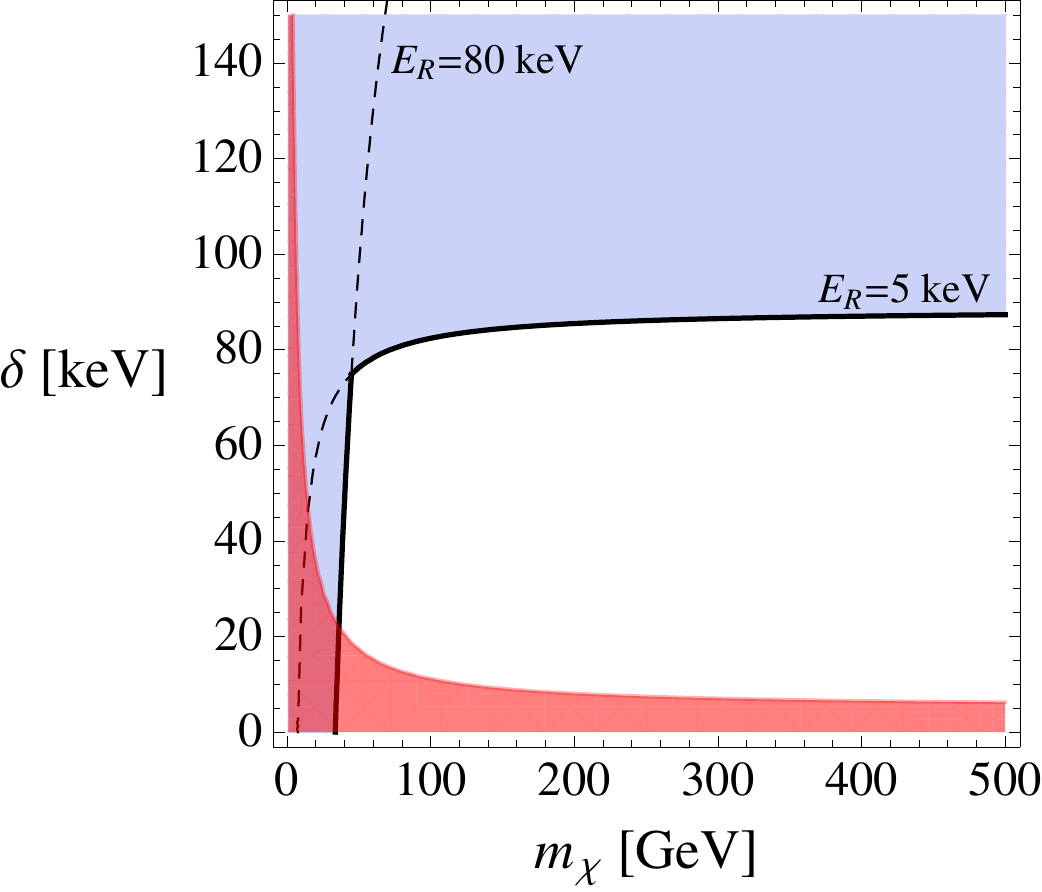}
\hspace*{0.05\textwidth}
\caption{The astrophysics independent allowed region of 
iDM parameter space (unshaded region) 
for a recoil spectrum with events, in a xenon detector assuming a 
maximal speed for dark matter of $500$~km/s (LH plot) and 
$800$~km/s (RH plot) in the Earth's frame.  
In both plots the two curves assume events observed with 
$5$ and $80$~keV\@.  The lower (red shaded) region is model independent 
and is the region in which the peak, Eq.~(\ref{eq:recoilpeak}), 
in the deconvoluted recoil spectrum lies below $5$~keV and iDM cannot 
be distinguished from eDM\@.  With reliable knowledge of $f_1(v)$ 
this region extends further upwards.}
\label{fig:iDMkinematics}
\end{center}
\end{figure*}

If $E_R^{\rm peak}$ is 
large enough there may even be an observable gap between an 
experiment's lower threshold and the first recoil events -- 
this bump-like spectrum aids in fitting the energy spectrum of the
DAMA modulation data.  On the other hand, if $E_R^{\rm peak}$ is below 
the lower threshold it is not possible to distinguish iDM from 
conventional elastic dark matter.  In Fig.~\ref{fig:iDMkinematics} 
we illustrate these points for the case of a xenon experiment which 
records data over the range $5\ \kev\le E_R\le 80\ \kev$.

Unlike eDM, where the recoil spectrum must be 
monotonically decreasing, the mass splitting of iDM allows
for the spectrum to be a rising function for $E_R < E_R^{\rm peak}$.  
This cannot be mimicked by elastic scattering without a dark matter 
form factor, no matter what (physically allowed) 
form the velocity distribution takes.  Furthermore, at $E_R^{\rm peak}$ 
the denominator in (\ref{eq:ffrominversion}) has a zero but since 
$f_1(v)$ must be finite the recoil spectrum must have a maximum 
at this same energy so the numerator goes to zero as well.   
Thus the observation of a maximum in the recoil spectrum determines 
$E_R^{\rm peak}$ and reduces the 3 dimensional iDM parameter space 
by one dimension, 
relating $m_\chi$ and $\delta$, independent of knowledge of the 
dark matter velocity distribution.

\subsection{Form Factor versus Inelastic}
\label{sec:ffidm}

Although the spectrum of iDM cannot be faked by simple 
elastically scattering DM, regardless of what physical 
velocity distribution is used, it may be faked by elastically scattering 
dark matter that has a non-trivial form factor.  
Due to the unique kinematics of iDM, induced by the splitting of the states, 
a deconvoluted iDM spectrum has a gap in the low energy spectrum 
with no recoil events.  The size of this gap, and thus the position 
of the first events is  
\be
E_R^{\rm gap} = \frac{\mu^2}{m_N} \left( 
  v_{\rm max}^2 - \frac{\delta}{\mu} 
  - \sqrt{\left(v_{\rm max}^2-\frac{\delta}{\mu}\right)^2 
  - \frac{\delta^2}{\mu^2}} \right)~.
\ee
This can be obtained directly from Eq.~(\ref{eq:ERsoln}) by
calculating the lowest recoil energy for the highest velocity 
($v = v_{\rm max}$) for a head-on collision 
($\cos\theta_{\rm lab} = 1$).
The spectrum then increases to $E_R^{\rm peak}$, Eq.~(\ref{eq:recoilpeak}), 
before again decreasing.  Near the peak, the spectrum is well fit 
by a power law but away from the peak the spectrum becomes 
more exponential in nature.  This increase in $\mathcal{R}$ 
between $E_R^{\rm gap}$ and $E_R^{\rm peak}$ would require an 
unphysical $f_1(v) < 0$ if the dark matter has no form factor.  
However, it is possible that a non-trivial $F_\chi(E_R)$ 
could ``overpower'' this increase and allow 
iDM spectra to be fit by eDM with a form factor, albeit for a 
non-Maxwellian velocity distribution.  We demonstrate this below 
for the simple example of a form factor that is a single power in 
$E_R$, $F_\chi\sim E_R^n$, for a sufficiently high $n$ the resulting 
$f_1(v)$ will be sensible.

Although the increase in the recoil spectrum can be accommodated 
by elastically scattering dark matter with a non-trivial form factor, 
the existence of a gap with iDM cannot be faked with a simple power-law
form factor.  This is because the threshold to up-scatter results 
in a step function turn on in the spectrum, and no such step function
results from a simple polynomial form for $F^2_\chi(E_R)$.

\begin{figure}[t]
\begin{center}
\includegraphics[width=0.40\textwidth]{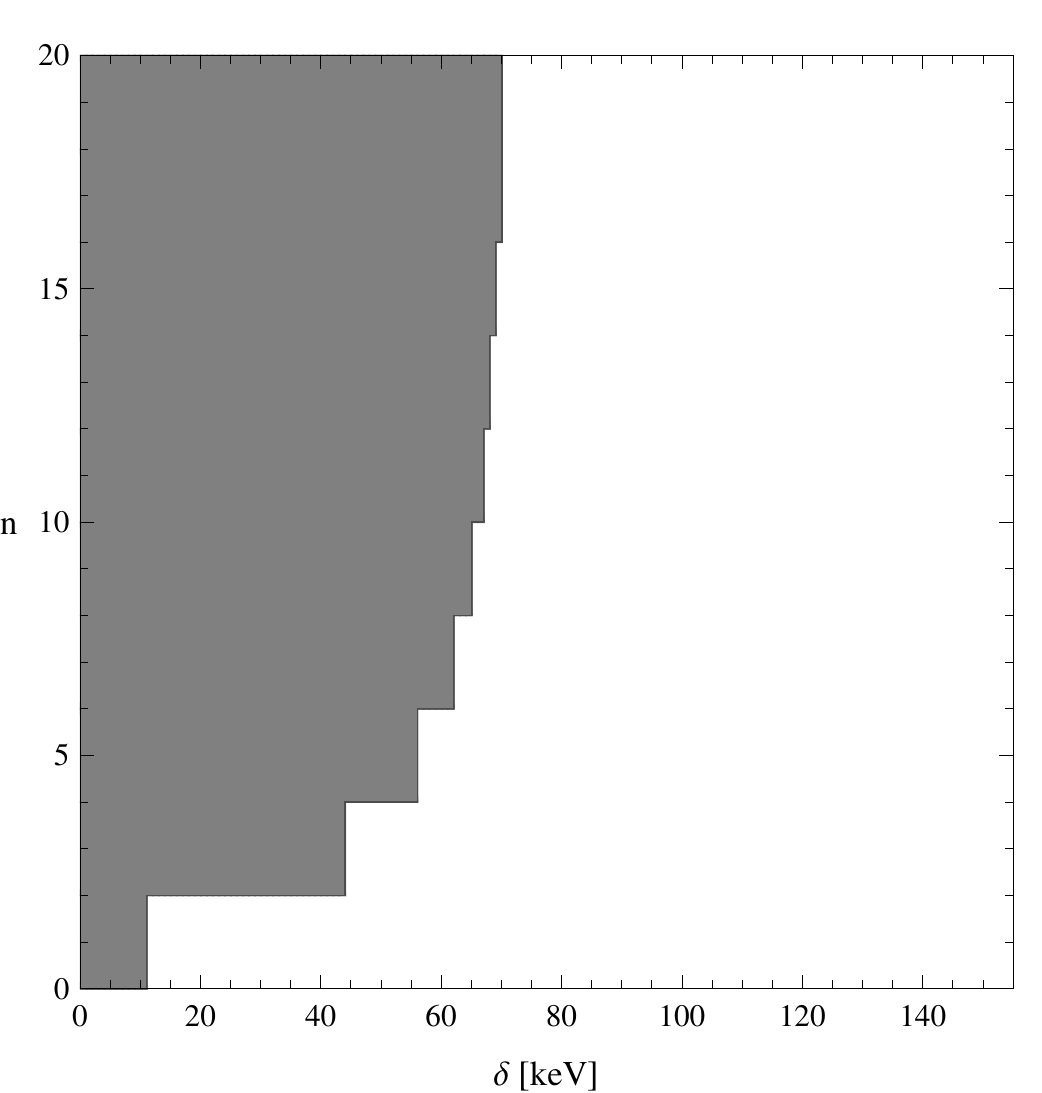}
\caption{In the shaded region elastically scattering dark matter with 
a form factor $F_\chi \propto q^{2n}$ can fake iDM with splitting $\delta$. 
The dark matter mass was taken to be $100$~GeV, scattering off xenon, and 
the iDM spectrum was assumed to come from a Maxwellian distribution.}
\label{fig:iDMfake}
\end{center}
\end{figure}

\begin{figure}[t]
\begin{center}
\includegraphics[width=0.49\textwidth]{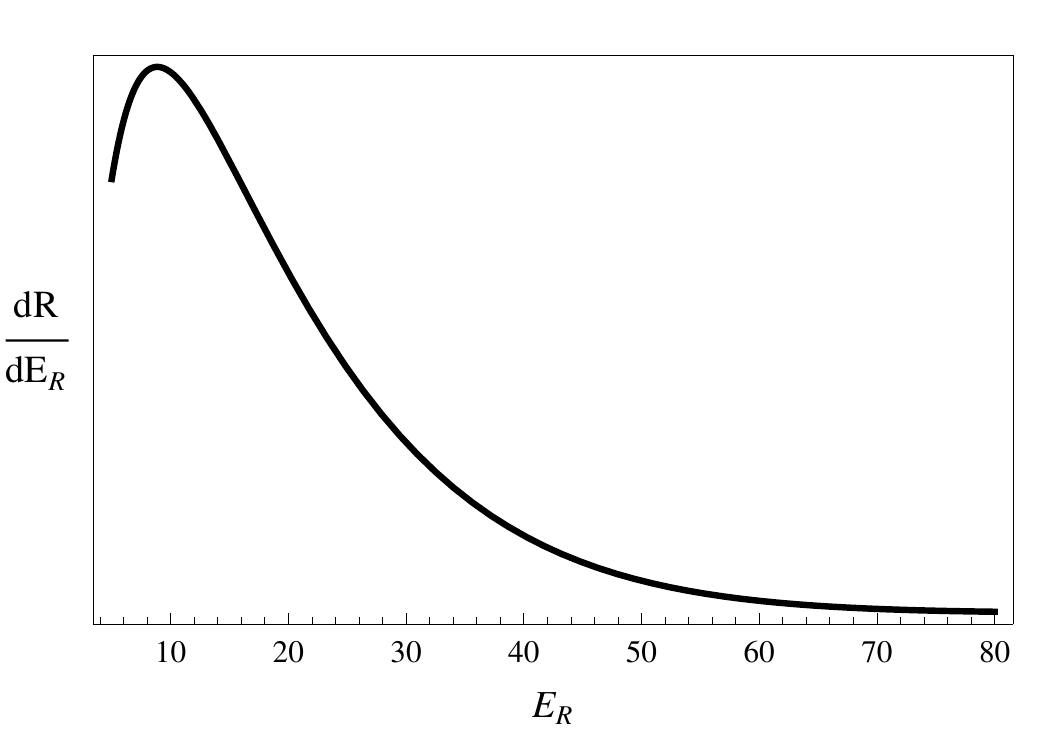}
\includegraphics[width=0.49\textwidth]{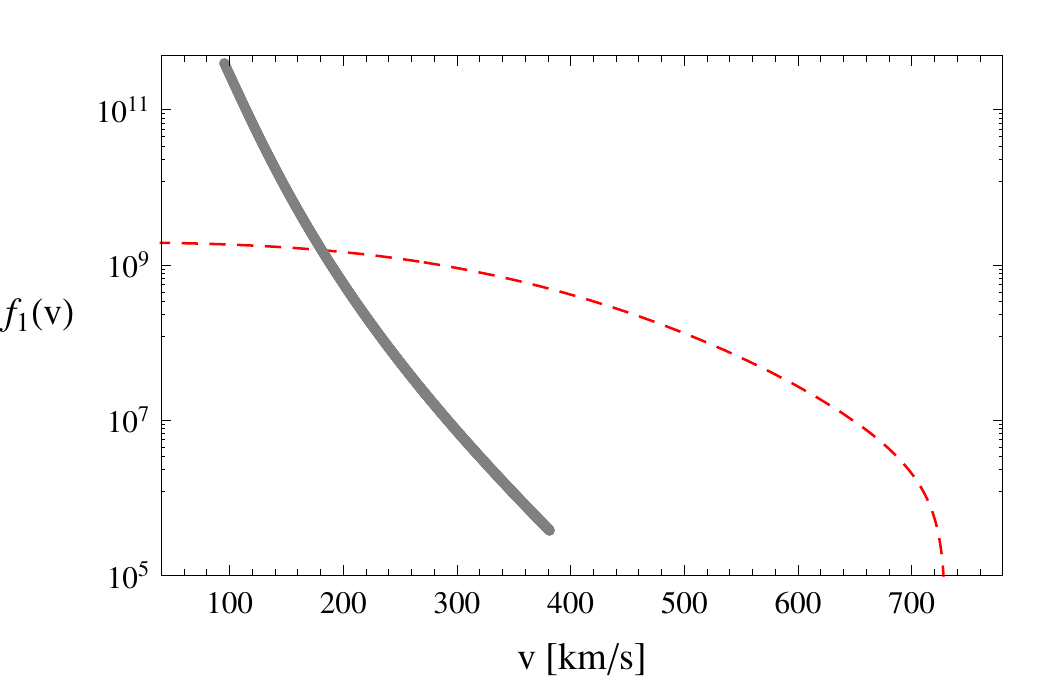}
\caption{Top: a recoil spectrum, at a xenon based detector, for iDM 
with $m_\chi=100$~GeV and $\delta= 35$~keV
within the range $5 \le E_R/\kev \le 80$.
Bottom: the iDM Maxwellian velocity distribution is shown as the 
the red dashed curve.  The solid gray curve corresponds to the
velocity distribution that exactly matches the recoil spectrum
shown in the top figure, but with dark matter, of the same mass, 
scattering elastically with a form factor $F_\chi\propto q^4$.}
\label{fig:iDMfakerveldisp}
\end{center}
\end{figure}

To illustrate the ability of iDM to be faked by form factor dark matter.  
We consider the simple case of $F_\chi = \left(\frac{q}{q_0}\right)^{2n}$ 
with the normalization $q_0= 10^{-3} \mu$ and we take $n\in  [-1,10]$ 
and investigate whether there is a physically sensible velocity distribution 
that allows form factor dark matter (with the same WIMP mass) to fake iDM with 
$m_\chi=100$~GeV, with a Maxwellian velocity distribution, 
for various splittings $\delta$.  This region in which iDM can indeed 
be faked is shown as the shaded region in Figure~\ref{fig:iDMfake} 
and an example velocity distribution that achieves this is shown 
in Fig.~\ref{fig:iDMfakerveldisp}.

\section{Discussion}

We have shown that certain particle physics properties
can be determined from the nuclear recoil spectrum
of a dark matter direct detection experiment
independent of the local astrophysical density and 
velocity distribution. 
These properties are most easily uncovered from the
deconvoluted scattering rate, that we call $\mathcal{R}$.
We advocate that experiments present their recoil spectrum
data in terms of the deconvoluted spectrum, from which
the shape of $\mathcal{R}$ as a function of $E_R$ can be 
read off independently of the nuclear form factor.

The kinematics of scattering leads to a lower bound on dark matter mass, 
given a maximum speed for the WIMP in the Earth's frame, 
independent of the details of the dark matter velocity distribution.
We have also shown that an upper bound on the mass of dark matter, however,
\emph{cannot} be determined independently of the 
density and velocity distributions of dark matter.
This was demonstrated for both ``fake'' Xenon data,
but also for the CoGeNT excess.  Perhaps the most striking
consequence is that we find the CoGeNT excess can be fit
by dark matter with masses much heavier than has been
generally considered.  Much of this parameter space,
up to $m_{\chi} \simeq m_{\rm Ge}$, is no more
strongly constrained than light dark matter, given an appropriate choice of
velocity distribution.
For $m_{\chi} > m_{\rm Ge}$, the CoGeNT data may or may not 
be allowed by existing Xenon10 and Xenon100 data, which 
ultimately depend on the details of $L_{\rm eff}(E_R)$ and $Q_y$.

Our analysis comparing different masses, as well as comparing
and contrasting inelastic dark matter with form factor dark matter,
was done without any experimental errors.  It would be interesting
to carry out similar analyses with experimental errors.
For example, the so-called ``gap'' in inelastic dark matter
may not be resolvable due to experimental errors, so that
it may be reproduced by a smooth dark matter form factor.

Breaking from the vast majority of past literature, 
we do not impose any requirements of the velocity distribution
of dark matter, except for assuming there is an upper bound
(that might or might not be equated to $v_{\rm earth} + v_{\rm esc}$
in Earth's frame).  In fact, our velocity distributions were
taken to be in Earth's frame, and thus to compare with most
other literature one would have to boost into galactic frame.

Interestingly, however, there are potential physical realizations  
of some of the more unusual velocity distributions found in this paper.  
In particular, distributions with velocities that are 
smaller than the Earth's velocity in the galactic frame 
($\simeq 230$~km/s) could be interpreted in at least two ways.
One is dark matter
that streams in a direction parallel to the motion of the 
Sun around the galaxy.  This could be a local stream, or could
be a ``dark disk''.  Another interpretation is that the lower
velocities arise from dark matter that is weakly bound to the
Earth/Sun/Jupiter system.  In both cases, relative velocities
in the tens, rather than hundreds, of km/s are physically
realizable.

There are still many questions and many interesting avenues
for future exploration of astrophysics-independent implications
of dark matter.  Time-dependence in the scattering rate is
perhaps the most important, given the ongoing interest
in the annual modulation observed by DAMA.  In the future,
directional dependence will also provide an excellent way to
determine the nature of the dark matter velocity distribution
we find ourselves in.  Understanding the particle physics
implications of these observations is left to future work.  
With information about dark matter from other sources, such 
as mass determinations from the LHC, the approach presented 
here could allow the interpretation of direct detection 
results as a direct measurement of the local astrophysical 
properties of dark matter.
Finally, we have not considered the implications of 
multiple nuclear recoil spectra from different experiments.
This too is an very interesting and timely subject that
we hope to see work on very soon~\cite{thatotherpaperbythatoneguyandhisminions}.

\section*{Acknowledgments}

We thank R.~Lang, R.~Fok, and L.~Strigari for useful discussions.  
We also thank the Kavli Institute of Theoretical Physics and the 
Aspen Center of Physics for each providing a simulating
atmosphere where part of this work was carried out.
GDK was supported by a Ben Lee Fellowship from Fermilab and 
in part by the NSF under contract PHY-0918108.
TMPT was supported in part by the NSF under contract PHY-0970171.
Fermilab is operated by Fermi Research Alliance, LLC, under 
Contract DE-AC02-07CH11359 with the United States Department of Energy.


\end{document}